\documentclass[sigconf,nonacm]{acmart}
\usepackage{multirow}
\usepackage{amsmath}
\usepackage[ruled,linesnumbered]{algorithm2e}
\usepackage{comment}
\usepackage{listings}
\usepackage{svg}
\usepackage{paralist}
\usepackage[inline]{enumitem}
\usepackage{xspace}
\usepackage{xcolor}
\usepackage{tcolorbox}
\usepackage{makecell}
\usepackage{tabularray}

\UseTblrLibrary{functional}

\newboolean{showcomments}
\setboolean{showcomments}{true}         
\ifthenelse{\boolean{showcomments}}
{\newcommand{\nb}[2]
    {
		\fbox{\bfseries\sffamily\scriptsize#1}
		{\sf\small$\blacktriangleright$\textit{#2}$\blacktriangleleft$}
	}
}
{\newcommand{\nb}[2]{}
	
}
\newcommand\giovanni[1]{\textcolor{red}{\nb{Giovanni}{#1}}}
\newcommand\luca[1]{\textcolor{cyan}{\nb{Luca}{#1}}} 
\newcommand\enea[1]{\textcolor{violet}{\nb{Enea}{#1}}}

\newcommand{\acronym}{\textsc{MCP}\xspace}
\newcommand{\nbcc}{\acronym-\textsc{Tester}\xspace}
\newcommand{\nabcc}{\nbcc}
\newcommand{\inputSrc}{seed input sources\xspace}
\newcommand{\baseInput}{base input source\xspace}

\newcommand{\numR}{10\xspace}

\newcommand{\confParamMRName}{\textsc{Add Configuration Parameter}\xspace}

\lstdefinelanguage{dsl}{
    morekeywords = [1]{ declarations, fsm, configure },
    morekeywords = [2]{ parameter, external, actuator, argument, configuration, command, variable },
    morekeywords = [3]{ if, else, true, false },
    morekeywords = [4]{ Boolean, Enumeration, seconds },
    morekeywords = [5]{ init, cycle },
    morekeywords = [6]{ transition, send, to, set, is },
    morekeywords = [7]{ state, pulse },
    keywordstyle = [1]\color{teal}\bfseries,
    keywordstyle = [2]\color{violet}\bfseries,
    keywordstyle = [3]\bfseries,
    keywordstyle = [4]\color{orange},
    keywordstyle = [5]\color{teal},
    keywordstyle = [6]\color{violet},
    keywordstyle = [7]\color{blue}
}
\lstdefinelanguage{diff}{
    morecomment = [l][\color{green!70!black}]{+},
    morecomment = [l][\color{red}]{-},
    morecomment = [s][\color{magenta!63!blue}]{@@}{@@}
}
\lstdefinelanguage{c2}{
    language = C,
    morekeywords = [1]{ once },
    morekeywords = [2]{ _Bool, trafficlight_t, trafficlight_state_t },
    morekeywords = [3]{ B_TRUE, B_FALSE, CAMERA_ACTCMD_ID, SPEAKER_ACTCMD_ID },
    morecomment = [s][\color{green!40!black}]{<}{>},
    morecomment = [l][\color{green!60!blue}]{//},
    commentstyle = \color{green!60!blue},
    keywordstyle = [1]\color{violet},
    keywordstyle = [2]\color{orange!65!black},
    keywordstyle = [3]\color{blue!70!black},
    stringstyle = \color{green!40!black}
}
\makeatletter

\let\origthelstnumber\thelstnumber
\newcommand*\lstspSuppressLineNumber{%
  \lst@AddToHook{OnNewLine}{%
    \let\thelstnumber\relax%
     \advance\c@lstnumber-\@ne\relax%
    }%
}

\newcommand*\lstspReactivateLineNumber[1]{%
  \setcounter{lstnumber}{\numexpr#1-1\relax}
  \lst@AddToHook{OnNewLine}{%
   \let\thelstnumber\origthelstnumber%
   \refstepcounter{lstnumber}%
  }%
}

\def\lst@PlaceNumber{\ifnum\value{lstnumber}=0\else
  \llap{\normalfont\lst@numberstyle{\thelstnumber}\kern\lst@numbersep}\fi}

\makeatother

\AtBeginDocument{%
  }

\begin{document}

\title{Metamorphic Testing of Transpilers via Mutation Consistency of Programs}

\author{Enea Raffaele Ilario Papaleo}
\orcid{0009-0000-9573-9657}
\email{e.papaleo1@campus.unimib.it}
\affiliation{%
  \institution{University of Milano-Bicocca}
  %\streetaddress{P.O. Box 1212}
  \city{Milano}
  \country{Italy}
}

\author{Luca Guglielmo}
\email{luca.guglielmo@unimib.it}
\orcid{0000-0002-5580-3623}
\affiliation{%
  \institution{University of Milano-Bicocca}
  %\streetaddress{P.O. Box 1212}
  \city{Milano}
  \country{Italy}
}

\author{Giovanni Denaro}
\orcid{0000-0002-7566-8051}
\email{giovanni.denaro@unimib.it}
\affiliation{%
  \institution{University of Milano-Bicocca}
  %\streetaddress{P.O. Box 1212}
  \city{Milano}
  \country{Italy}
}

\renewcommand{\shortauthors}{Papaleo et al.}

\begin{abstract}
Transpilers are increasingly used for software development, especially in industrial domains that rely on domain‑specific languages (DSLs), to allow engineers to work with familiar concepts and appropriate abstractions. Ensuring the correctness of these instruments is therefore critical in many industrial settings. This paper observes that existing approaches for compiler testing hardly generalize to transpilers. Differential testing approaches are hindered as multiple equivalent implementations of the transpiler under test are seldom available in practice. The approaches based on metamorphic testing assume the ability to execute the compiled binaries, an assumption that cannot be always made for transpilers, which oftentimes produce results expressed as 
source code, requiring complex toolchains,  hardware‑in‑the‑loop setups, and depending on non trivial inputs.
This paper introduces a novel metamorphic testing technique tailored to transpilers. Instead of reasoning about the runtime behavior of compiled programs, our approach defines metamorphic relations directly over the source code produced by the transpiler. These relations capture a property that we call mutation consistency of the (transpiled) programs: mutation‑style changes in the input DSL program must induce predictable and structurally consistent changes in the generated output. We implemented this idea in a tool, \nbcc, and evaluated it through a case study conducted in the context of a  technology‑transfer project.
Our current empirical results indicate that the proposed approach can effectively reveal faults that would remain undetected with pure fuzzing. 
%The study demonstrates that metamorphic testing based on mutation consistency provides a practical and effective solution for validating transpilers in real‑world settings.
\end{abstract}





\maketitle

\section{Introduction}
\label{sec:intro}
%\luca{1) I transpiler sono un dominio importante. 2) Riassunto sezione 2.2. Ci sono soluzioni, ma per il nostro caso non sono applicabili. 3) Dire che il nostro approccio è nuovo perchè risolve gli open problem precedenti. 4) Una breve menzione ai dati empirici che dimostrano l'efficacia delle tecnica. 5) Contenuto del paper.}

%1) I transpiler sono un dominio importante
Transpilers, also known as source‑to‑source compilers, play a central role in modern industrial software development. A transpiler translates programs written in a source language into code expressed in another (target) programming language, enabling engineers to work with higher-level or domain-specific concepts, while obtaining code that is more suitable  for compilation and execution~\cite{bastidas2023transpilers,hirzel2016code,chaber2016effectiveness}.
%Several transpilers have been developed across the years and their domain of application are quite varied, both in research and industry. One of earliest transpilers dates back to the 70s~\cite{MCS-86} and was used to convert 8-bit programs is 16-bit ones. In recent years, more widely known transpilers are the ones used to convert more common programming languages, like C, into other ones, such as Rust~\cite{ling2022rust, leopoldseder2015java, marcelino2022transpiling, typescript, doeraene2013scala}.
Their relevance has grown significantly with the widespread adoption of domain‑specific languages (DSLs)~\cite{kosar2016domain,wkasowski2023domain,hudak1997domain}. DSLs allow domain experts to express system behavior using familiar concepts and terminology, but the resulting specifications typically require translation into a general‑purpose programming language before they can be compiled or deployed. As a consequence, the correctness of transpilers becomes mission‑critical: faults in the translation process may silently propagate into the generated code and ultimately into deployed systems.
%To validate the correctness of the transpilation process, testing is essential. In particular, testing activities must ensure that no fault in the transpiler affects its output resulting in erroneous code being generated. Similarly to other software contexts, the high costs of manual testing must be considered, and consequently the usage of test generation and automation should be evaluated. 

%2) Riassunto sezione 2.2. Ci sono soluzioni, ma per il nostro caso non sono applicabili.
%The literature on generating test cases for transpilers is practically non-existent, but researchers can take inspiration from the one on compiler testing. In particular, the literature highlights two main approaches for effectively testing compilers: differential testing and metamorphic testing~\cite{chen2020survey}. Differential testing requires the existence of multiple, supposedly equivalent implementations of the compiler, in order to cross-check the results obtained while testing a compiler against the results of an equivalent counterpart. On the other hand, metamorphic testing relies on metamorphic relations, which specify how particular changes to the test input are expected to change the output.
Despite their importance, automated testing techniques for transpilers remain largely unexplored. A possible workaround is to borrow solutions devised for compiler testing, drawing on the assumption that a transpiler is a "sort of" compiler.
Existing research on compiler testing offers two main families of approaches: differential testing and metamorphic testing.
%\cite{chen2020survey}. 
Differential testing compares the outputs of multiple, supposedly equivalent compiler implementations
\cite{ofenbeck2016randir,sun2016finding2,hawblitzel2013will,li2023finding}, but is inapplicable when only a single implementation exists, as it is common for many industrial transpilers. Metamorphic testing for compilers~\cite{le2014compiler,le2015finding,tao2010automatic,nakamura2016random,donaldson2017automated,chen2020survey} defines relations over source programs whose compiled binaries should behave equivalently at runtime, exploiting the expected equivalences as automated test oracles. These techniques draw on a fundamental assumption: the compiler produces executable binaries that can be run with available or easily derivable inputs. However, this assumption often breaks down in the case of transpilers, as their output is itself source code, often requiring complex toolchains, hardware‑in‑the‑loop setups, or domain‑specific simulators for compilation and execution. Moreover, %while also 
the mapping of the input space across the original and the transpiled sources can be far than trivial.
In such settings, existing compiler‑oriented metamorphic testing approaches may become ineffective or even inapplicable.
%On the other hand, current works on metamorphic testing of compilers have the potential to be more relevant for our uses. These works rely on metamorphic relations that either exploit transformations of non-executed code regions (dead code) of the test programs~\cite{le2014compiler,le2015finding}, or transformation rules to obtain test programs yielding equivalent outputs~\cite{tao2010automatic,nakamura2016random,donaldson2017automated}.
%Unfortunately even these techniques have their limitations. In particular, they can be ineffective in situations in which a transpiler bug affects in equivalent way two input programs.
%both the original program and the transformed one. 
%In this scenario, the resulting binaries will exhibit the same failures, preventing the metamorphic oracles from detecting behavioral differences and therefore making the underlying fault unobservable.

%3) Dire che il nostro approccio è nuovo perchè risolve gli open problem precedenti
This paper introduces a novel metamorphic testing technique specifically designed for transpilers. Instead of reasoning about the runtime behavior of compiled binaries, our approach embraces metamorphic relations defined directly over the source code produced by the transpiler. 
These metamorphic relations specify how \emph{mutation-style changes in the input source code} are expected to \emph{consistently map onto corresponding equivalencies and differences in the output source code} produced by the transpiler under test. 
We call the underlying property Mutation Consistency of transpiled Programs (\acronym). This shift from runtime equivalence of compiled binaries to structural consistency of generated source code enables metamorphic testing in contexts where traditional compiler‑testing techniques cannot operate.

%While these testing techniques are sound, they do not generalize to all possible contexts. In particular, through the collaboration with an industrial partner, we identified the need to test a transpiler with characteristics that make the usage of existing testing techniques impractical, or outright impossible. In our case study, differential testing cannot be applied, since we do not have two different and supposedly equivalent implementations of the transpiler.
This idea emerged from a technology‑transfer project with an industrial partner, whose transpiler translates control‑logic DSL programs into executable code for industrial automation systems. This collaboration provided both the motivation and the opportunity to evaluate our technique in a realistic setting. So far, we conducted a first case study in which we applied the \acronym metamorphic testing for validating the partner’s transpiler. The case study also provided the ground for the experiment that we report in this paper, in which we 
injected faults into the bytecode of the partner’s transpiler 
and assessed the ability of our approach to reveal those faults. The empirical results indicate that the \acronym approach can effectively detect faults that would
remain unexposed with pure fuzzing, testifying the practical significance and the potential of the approach proposed in this paper.
%4) Una breve menzione ai dati empirici che dimostrano l'efficacia delle tecnica.
%We implemented and investigated the effectiveness of \acronym in the context of an empirical study which is part of a technology-transfer project joint with an industrial partner. The results demonstrate the effectiveness of our approach in detecting potential faults in the transpiler, outperforming a baseline that relies only on transpiler fuzzing and is therefore limited to uncovering only crash-inducing faults.

%5) Contenuto del paper.
The remainder of the paper is organized as follows. Section~\ref{sec:example} reviews existing approaches to compiler testing and highlights why they do not necessarily generalize to transpilers. Section~\ref{sec:mtt} introduces the \acronym approach and presents \nbcc, an automated test generator based on the \acronym approach. Section~\ref{sec:casestudy} reports our empirical evaluation conducted during the case study with the industrial partner. Section~\ref{sec:related} discusses related work.  Section~\ref{sec:conclusions} summarizes our contribution and future research directions.

\section{State of the Art and Overview of the Proposed Approach}
\label{sec:example}
In this section, we survey the main approaches to applying metamorphic testing for
compiler testing, acknowledging that these approaches can also be used to test
transpilers. We center the discussion around a sample bug of a flawed transpiler
which reveals the limitations of the existing approaches. We then refer to the same
example to highlight the distinctive characteristics of the approach that we propose
in this paper. 

%In this section, we will show an example of usage of a transpiler for the conversion of a custom domain-specific language (DSL) into C code. We will then discuss thelimitations that existing metamorphic testing techniques for compilers have when applied to the case currently being examined. Finally, we will describe the  original contributions of the technique outlined in this paper, applying it to the same example.

\subsection{Working Example}
\label{sec:working:example}
As a working example, we introduce a sample transpiler, which aims to compile source programs written in a custom domain-specific language (DSL) into corresponding C code.
We consider a DSL that mimics (modulo adaptations and simplifications for the purpose of making the presentation accessible) a DSL defined by an industrial partner. This language is designed to support their domain analysts in specifying the control logic of industrial plants by means of a state-machine-based formalism, that the analysts (who are not professional IT programmers) can use more effectively than conventional programming languages.
%can manage more comfortably than a classic programming language.
They can then use the transpiler to translate the DSL control logics into actual C programs, which can be compiled and  deployed.

\begin{figure}
\begin{lstlisting}[
    language = dsl,
    basicstyle = \scriptsize
]
declarations {|\label{lstln:semdsl:decbegin}|
  variable state: Enumeration { Red, Green, Yellow }|\label{lstln:semdsl:state}|
  actuator command camera(argument enable: Boolean)|\label{lstln:semdsl:camact}|
  actuator command speaker(argument enable: Boolean)|\label{lstln:semdsl:spkact}|
  configuration parameter enableCamera: Boolean|\label{lstln:semdsl:camparam}|
}|\label{lstln:semdsl:decend}|
fsm {|\label{lstln:semdsl:fsmbegin}|
  init { transition to Red }|\label{lstln:semdsl:fsminit}|
  cycle {|\label{lstln:semdsl:fsmcyclebegin}|
    if (enableCamera) {|\label{lstln:semdsl:camcheck}|
      if (state is Red) send command camera(enable = true)|\label{lstln:semdsl:camon}|
      else send command camera(enable = false)|\label{lstln:semdsl:camoff}|
    }
    send command speaker(enable = state is Red)|\label{lstln:semdsl:spkset}|
    transition to state.next|\label{lstln:semdsl:nexttrans}|
  }|\label{lstln:semdsl:fsmcycleend}|
}|\label{lstln:semdsl:fsmend}|
configure {|\label{lstln:semdsl:cfgbegin}|
    set parameter enableCamera to false|\label{lstln:semdsl:camcfg}|
    set pulse to 10 seconds|\label{lstln:semdsl:cfgpulse}|
}|\label{lstln:semdsl:cfgend}|
\end{lstlisting}
\caption{A sample traffic-light controller in a custom DSL}
\label{lst:semdsl}
\end{figure}

Figure~\ref{lst:semdsl} shows a sample logic written in the considered DSL.
It defines a finite-state machine that represents the behavior of a simple traffic light.
%, which we use as working example. 
%custom DSL used to 
%As can be inferred from the provided example, the DSL has been designed for usage by non-programmers, as the syntax is similar to the natural English language, albeit with more structured elements. The logic is divided into three sections, which we will explore in order.
%
%In 
The program begins with the declarations of
%(lines~\ref{lstln:semdsl:decbegin}\textendash\ref{lstln:semdsl:decend}). %lies the declaration area, which is where elements that are necessary for the correctfunctioning of the state machine are defined. 
%In particular, 
%is the most important as it 
%defines 
the set of states that
the traffic light can assume (line~\ref{lstln:semdsl:state}),
%, named after the traditional three colors:
%in this case one for each of its lights. These states
%are also ordered based on the order of their declaration, which matches the usual
%intuition. Some traffic lights are also equipped with a camera that is usually
%toggled on or off depending on the light that is currently illuminated, in order to
%identify drivers running a red light. Line~\ref{lstln:semdsl:camact} models this
%component, specifying an external actuator that can be enabled or disabled via the
%use of commands. In a real-life scenario, this command could be wired into some
%circuitry that can enable or disable the ability of the camera to snap pictures. Since
%not all traffic lights come with a camera, though, the logic must be able to avoid
%sending commands to a non-existent device, as this could potentially lead to a stall
%while waiting for an answer. 
%For this reason, 
%Line~\ref{lstln:semdsl:camparam} introduces a parameter, which can be set to true to activate a camera. 
%represents a configuration variable that gets set by the operator prior to the execution of the logic. 
the relevant actuator commands (line~\ref{lstln:semdsl:camact} and line~\ref{lstln:semdsl:spkact}), 
%i.e., a camera device and a speaker device, respectively;
and
%introduces 
a configuration parameter that
can be set to activate or deactivate the camera (line~\ref{lstln:semdsl:camparam}). 
%models the component of a speaker that emits a sound whenever the traffic light is red, to signal to sight-impaired pedestrians that they may cross the road, in a similar way to the camera. Differently from the camera, though, due to accessibility laws in the area the traffic lights are to be deployed in, the speaker is mandatory.For this reason, no configuration parameter for it is present.
Next, lines~\ref{lstln:semdsl:fsmbegin}\textendash\ref{lstln:semdsl:fsmend} contain the
core of the state machine logic. Like in most embedded software programming, logic
is split into two phases, the initialization phase and the cycle phase, with the former
being executed only once on power-on, and the latter looping at fixed time intervals
until the machine loses power. In particular, line~\ref{lstln:semdsl:fsminit}
specifies the initialization logic that must be executed on power-on to properly
set up the state machine. In this case, the logic ensures that the traffic light activates its red light at power-on.
The \texttt{transition to} instruction is used to instruct the internal logic to queue up a transition to the specified state at the end of the cycle.
The cycle logic is located at lines~
\ref{lstln:semdsl:fsmcyclebegin}\textendash\ref{lstln:semdsl:fsmcycleend}.
Initially, the presence of a camera for this specific traffic light installation
is checked (line~\ref{lstln:semdsl:camcheck}), so that the proper enabling or
disabling logic can be carried out: if the traffic light is currently showing a
red light, then the camera must be enabled, which is accomplished by the use
of the \texttt{send command} construct as shown on line~\ref{lstln:semdsl:camon};
on the other hand, the camera must be disabled in all other situations, thus
line~\ref{lstln:semdsl:camoff} sends the opposite command as needed. Following this, the
traffic light must illuminate the next light in the sequence\textendash which we
assume to be the usual red-green-yellow cycle\textendash and enable or disable
the pedestrian speaker accordingly. The code at lines~
\ref{lstln:semdsl:spkset}\textendash\ref{lstln:semdsl:nexttrans} accomplishes
this behavior by leveraging the ordering of the states previously
declared at line~\ref{lstln:semdsl:state}.
%in the \texttt{declarations} section on
%We want to underline how this DSL does not provide wrap-around mechanics,meaning that the special case in which the green light must be illuminated afterthe red one turns off needs to be handled explicitly, which this code does atline~\ref{lstln:semdsl:redtrans}. At the same time, we take the time to enableor disable the speaker according to the specifications we outlined at the beginning of the example, thanks to the instructions at lines~\ref{lstln:semdsl:spkon}
%and \ref{lstln:semdsl:spkoff}.
%
Lines \ref{lstln:semdsl:cfgbegin}\textendash\ref{lstln:semdsl:cfgend} indicate the actual configuration. %section located at lines~\ref{lstln:semdsl:cfgbegin} \textendash\ref{lstln:semdsl:cfgend} is the area dedicated to the personalization of the logic for the specific case in which it must execute. In this case, 
Line~\ref{lstln:semdsl:camcfg} indicates that this particular traffic light is not
equipped with a camera.
%to snap photos to drivers running red lights, and that 
Line~\ref{lstln:semdsl:cfgpulse} sets the time interval between successive executions
of the loop to ten seconds.

\begin{figure}
\begin{lstlisting}[
    language = c2,
    basicstyle = \scriptsize,
    firstnumber = 0
]
|\begin{center}\small{\textbf{trafficlight.h}}\end{center}|
#pragma once
#include <stdbool.h>
#include "fsm.h"
typedef struct {|\label{lstln:semcbug:parstructbegin}|
  _Bool p_enableCamera;
} trafficlight_t;
typedef enum { // Bugged|\label{lstln:semcbug:bugbegin}|
  Yellow = 0,
  Green,
  Red,
} trafficlight_state_t;|\label{lstln:semcbug:bugend}|
void trafficlight_fsm_init(trafficlight_t *);
void trafficlight_fsm_cycle(trafficlight_t *, trafficlight_state_t);
_Bool get_enableCamera(trafficlight_t *);|\label{lstln:semcbug:parget}\lstspSuppressLineNumber|
|\begin{center}\small{\textbf{trafficlight.c}}\end{center}||\lstspReactivateLineNumber{1}|
#include "actuator_command.h"
#include "actuator_command_ids.h"
#include "fsm.h"
#include "trafficlight.h"|\label{lstln:semcbug:h}|
void trafficlight_fsm_init(trafficlight_t *instance) {
  fsm_transition_to(Red);
}
void trafficlight_fsm_cycle(trafficlight_t *instance,
                            trafficlight_state_t current_state) {
  if (get_enableCamera(instance)) {|\label{lstln:semcbug:deadbegin}|
    if ((current_state == Red) ? B_TRUE : B_FALSE) {
      send_actuator_command(CAMERA_ACTCMD_ID, B_TRUE);
    } else {
      send_actuator_command(CAMERA_ACTCMD_ID, B_FALSE);
    }
  }|\label{lstln:semcbug:deadend}|
  send_actuator_command(SPEAKER_ACTCMD_ID, 
    (current_state == Red) ? B_TRUE : B_FALSE);
  fsm_transition_to((current_state + 1) % 3);
}
_Bool get_enableCamera(trafficlight_t *instance) {|\label{lstln:semcbug:pargetc}|
  return instance->p_enableCamera;
}
\end{lstlisting}
\caption{The output of the execution of a (bugged) transpiler on the code in Figure~\ref{lst:semdsl}}
\label{lst:semcbug}
\end{figure}

The transpiler translates the DSL definitions, like the ones above, into
an equivalent program in C. Figure~\ref{lst:semcbug} shows the
translation produced for the logic defined in Figure~\ref{lst:semdsl}. 
%The result of the translation is provided in Figure~\ref{lst:semc}, which aims to be illustrative rather than completely accurate. 
%Note 
%the translation of %the configuration parameter with a \texttt{define} preprocessor instruction on line~\ref{lstln:semc:camdef}, and
%the translation of the possible states the finite state automaton can assume as an \texttt{enum} located at lines~\ref{lstln:semc:statebegin}\textendash \ref{lstln:semc:stateend}.
%
In our example, we assume that the transpiler has a fault that causes the entries in
an enumeration to be translated in the opposite order with respect to how they were
declared. In this case, the bug manifests itself between
lines~\ref{lstln:semcbug:bugbegin}\textendash\ref{lstln:semcbug:bugend} of
\texttt{trafficlight.h} in Figure~\ref{lst:semcbug}. The resulting code is still compilable
and executable, but
its behavior is now no more compatible with the logic encoded by the DSL. In fact, the
error makes the traffic light cycle according to a red-yellow-green cycle, rather than the
one specified in the DSL.

The code is still compilable and executable, but it is no more compliant with the DSL logic.
Infact, the behavior of the traffic light itself now erroneusly but consistently cycles
through the lights according to a red-yellow-green cycle.

\subsection{Current Techniques for Metamorphic Testing of Compilers \label{existing:techniques}}

At the state of the art, one can address the problem of generating test cases for
a transpiler by leveraging existing research on compiler testing. A complete approach
shall encompass the two main tasks of
\begin{inparaenum}[(i)]
\item generating test programs for exercising the compiler under test, and 
\item defining oracles for checking the results of the generated test cases~\cite{chen2020survey}.
\end{inparaenum}
The solution for the former task can be engineered 
%is relatively easier than the latter one, as  the space of possible inputs is precisely defined 
by grounding on the grammar of the input language (e.g., the DSL of our example).
%, we can generate test programs with several existing approaches;
%In general, as  the space of possible inputs is precisely defined by the grammar of the input language (e.g., the DSL of our example), we can generate test programs with several existing approaches; 
In contrast, identifying test oracles is most often a major challenge, due to both
missing a specification of the behavior of the target compiler and the hardness
of coping with the semantic richness of the input and output languages.

Researchers have explored 
%The recent survey of Chen at al.\ classifies 
two main approaches for defining oracles while generating test cases for
compilers: %based on either 
differential testing and metamorphic testing~\cite{chen2020survey}. Differential
testing requires the existence of multiple, supposedly equivalent implementations of the
compiler, in order to cross-check the results obtained while testing a compiler against
the results of an equivalent counterpart. In this paper, we aim at addressing scenarios
in which equivalent implementations are not available, and thus we consider solutions
based on differential testing out of our scope. On the other hand, metamorphic testing
relies on metamorphic relations, which specify how particular changes to the test input
are expected to change the output.

%Current metamorphic testing techniques focus their efforts on testing compilers rather than transpilers. Nevertheless, their insights can be used to on the latter category of software, as transpilers are very similar in architecture to compilers. In fact, their major difference mostly resides in the type of output, with compilers focusing on producing low-level machine code and transpilers on high-level source code for a specific programming language.

%According to a recent survey on compiler testing \enea{REF}, 

So far, researchers have proposed metamorphic testing of compilers grounded
on metamorphic relations
that exploit either non-executed code regions (dead code) of the test programs, or
transformation rules to obtain test programs yielding equivalent outputs.

\paragraph{Metamorphic Relations via Dead Code Manipulation}
A possible approach is to manipulate a test program is by modifying a region of the code
that is never executed. In this way, we obtain a second test program that, by
construction, should yield the same output as the original test program when they are
both executed with the same input. Thus, we can define a (metamorphic) test oracle by
checking that the two executable programs, obtained through compilation of the test
programs we previously obtained, actually exhibit the same behavior when executed with
a given set of inputs. Should there be a bug in the compiler under test, we expect that
this type of test oracle may fail for some possible dead code manipulations.
%
%One of the approaches presented sees the manipulation of dead code as a way to identify bugs during compilation, be either through removal or addition of dead code. Through manual analysis of the code, we can see that the logic in Figure~\ref{lst:semdsl} represented by the \texttt{if} statement at line~\ref{lstln:semdsl:camcheck} and its body are dead code.
%
%the technique of Le et al. \enea{REF: Le et al.}, 
Existing techniques are focused on either dead code elimination or code insertion in
dead regions~\cite{le2014compiler,le2015finding}. 
%we can remove portions of the code contained in this area,
%which will affect the output of the resulting translation. 

For example, with reference to the sample test program outlined in Figure~
\ref{lst:semdsl}, the test generator can notice that the logic comprised by the
\texttt{if} statement at line~\ref{lstln:semdsl:camcheck} is dead code, as variable
\texttt{enableCamera} is configured with the value \texttt{false}. Thus, it can
straightforwardly obtain equivalent test programs, by removing or adding statements
within the body of the \texttt{if} statement. Then, it can run the transpiler of the
example against a pair of equivalent test programs, turn the resulting C code into
corresponding binaries, and execute those binaries to check the equivalence of their
outputs. The test fails upon observing any difference in the output, evidence that the
transpiler under test produced non-equivalent C programs out of equivalent DSL sources.
Otherwise, the test passes and the test generator can iterate the process for other
pairs of equivalent test programs, while also potentially considering further
(automatically generated) test programs other than the one in Figure~\ref{lst:semdsl}.

However, we observe that this approach is unable to detect the fault of the transpiler
we have exemplified above. Even though the C code generated by the transpiler for both
our sample test program and the equivalent test programs created by the test generator
are all affected by the bug, all the corresponding binaries behave the same (and fail equivalently) at runtime. Therefore, metamorphic oracles based on dead code
manipulation will report a "\emph{test passed}" verdict in all cases. 

%the behavior of the C programs remain equivalent under 

%no line in the corresponding dead code fragment in Figure~\ref{lst:semcbug} (which is located at lines~\ref{lstln:semcbug:deadbegin}\textendash\ref{lstln:semcbug:deadend}) can result in detection of the bug. In fact, there is no reference to the now-missing symbol \texttt{Yellow} and removal processes cannot add them.

%A similar technique proposed by Sun et al. \enea{REF} relies on the insertion of code in both live and dead regions of the program. While the process they outline \textit{could} result in the addition of code that references the \texttt{Yellow} state and thus identifying the compiler error, we argue this is not particularly relevant for our approach. 
%In fact,  the mentioned technique  is centered around generating code that is semantic-preserving, to then ensure the behavior of the software after compilation is the same. %While the process of code generation may \textit{randomly} generate non-compilable code due to compiler faults, this is incidental. 
%In other words, a testing campaign executed to verify compiler faults of the mentioned type using the mentioned technique may or may not successfully identify the fault depending on what kind of code gets randomly generated during the insertion phase.

\paragraph{Metamorphic Relations via Equivalent Expressions, Assignments, and Submodules}
Other authors construct equivalent test programs by manipulating an initial test program
according to semantic-preserving program transformations. Tao et al.\@ obtain
%presents another approach to constructing %, based around construction of
equivalent programs based on %that are equivalent,
%. They defined
%This is done through the usage of 
%equivalence
equations that provide basic building blocks to construct equivalent expressions, which
in turn allow for building equivalent assignments and submodules~\cite{tao2010automatic}.
Similarly, it is possible to inject into the test programs suitably synthesized
semantic-preserving code~\cite{nakamura2016random}, or reorder some statements in the code,
or even apply other types of semantic-preserving program
transformations~\cite{donaldson2017automated}. 

%In the case of our DSL, we lack the ability to perform arithmetic operations, which mirrors the design of some real-world examples of DSLs for safety critical software. Therefore, the constructions of equivalent expressions as outlined in the paper, which is based around arithmetic tauotologies, cannot be carried out.

%As for the generation of equivalent assignment blocks, the method proposed in the paper relies on building an assignment order dependency graph and reordering assignments based on the graph. To simplify, assignment statements that are order-agnostic between themselves may be swapped. In the code of Figure~\ref{lst:semdsl}, the blocks containing more than one assignment statements are located at lines~\ref{lstln:semdsl:spkon}\textendash\ref{lstln:semdsl:redtrans} and lines~\ref{lstln:semdsl:spkoff}\textendash\ref{lstln:semdsl:prevtrans}. While neither of these are actual assignment statements in the traditional sense, we can interpret them as the equivalent of assignments for our DSLs. 

For instance, with reference to our sample test program, the test generator can detect
that the two last statements in the \texttt{cycle} block for the \texttt{fsm} in the
test program are order independent: changing the light of the traffic light and enabling
or disabling the speaker are two actions that can be carried out in any order.
%with respect to each other. 
%Thusly, an application of the technique by Tao et al. to this DSL could swap any of the two statements with the other in either of the two blocks. 
It can thus obtain an equivalent test program by swapping the two statements with each
other, execute the transpiler against both equivalent test programs, and verify the oracle
that the binaries compiled from the corresponding C code produce the same output at runtime.

Nevertheless, this approach is also unable to detect the fault of the transpiler, just as
the one we analyzed previously. In fact, the binaries obtained from the equivalent programs
still behave the same (and fail equivalently) at runtime.

%Nevertheless, 
%we can observe that this approach is unable to detect the fault of the transpiler, as the binaries obtained from the equivalent programs still behave the same (and fail equivalently) at runtime.
%neither of the two lines references the \texttt{Yellow} state by name and swapping lines cannot alter their contents. Therefore, this stage would not be able to identify the fault in the compiler.
%Finally, the used technique leverages classical optimization techniques such as dead code elimination to obtain equivalent submodules. We have already tackled the problem of dead code elimination in the previous section, showing how it is insufficient to identify the compiler bug we previously mentioned.

\subsection{Overview of Our Technique}
\label{sec:example:our}
In this paper, we propose a novel technique for testing transpilers, which exploits
metamorphic testing in a radically different way from the existing approaches on
compiler testing. We ground on the observation that a transpiler is distinctively
different from a traditional compiler, as it is not designed to produce an executable
binary; rather, it renders the given program to the corresponding source code. For
instance, the transpiler of our working example renders programs written in the DSL of Figure~\ref{lst:semdsl} to the corresponding source code in C, i.e. the C code of Figure~\ref{lst:semcbug}. This fundamental difference introduces technical issues that
do not necessarily show up in compiler testing, but also brings new opportunities.

On one hand, executing the programs produced by a transpiler can be less
straightforward than in the case of the binaries yielded by a compiler. For example,
executing the C code produced by the transpiler of our example requires the test generator
to first compile those C programs. This in turn may require the resolution of any
deployment-environment-related runtime dependencies, including dependencies that may be
outside the scope of
%out of order \enea{out of order??} with respect to
the specific problem domain of the
transpiler.
%, thus challenging the test generator to be specifically engineered with respect to the  runtime dependencies of each specific transpiler under test.  
Once the test generator succeeds in compiling the transpiled programs into binaries,
it must then identify suitable inputs to execute those binaries. The compiler testing
approaches surveyed above often rely on generating random inputs for this purpose. A
technical difference is that, in transpiler testing, the test generator shall elicit
inputs that fit the declarations in the output sources (the C code of our example) and
the deploying environment, rather than address directly the input space of the
original sources (the DSL programs of our example).

On the other hand, as the output of a transpiler is source code written in a
programming language, we see the opportunity to define oracles that predicate directly on
the textual representation of the output code. In the solution proposed in this paper,
we abandon the approach of evaluating the test cases by executing the binaries of the
semantically-equivalent programs, and embrace the idea of defining metamorphic relations
that specify how changes in the input sources are expected to map into observable
differences in the output sources. 
% by specifically considering the problem of transpiler testing,

We exemplify our approach with reference to the working example. A possible metamorphic
relation of the type defined in our approach could be the following:

\fbox{\parbox[t]{0.9\columnwidth}{
    \vspace{1pt}
    Let $P'$ and $P''$ be two test programs that differ only in the declaration of
    a single enumerative type $ENUM$, such that $ENUM$ is declared in both programs,
    but it includes an additional value identifier $NEWVAL$ in $P''$.
    
    Then, after
    transpiling $P'$ and $P''$, we expect that the output sources $O'$ and $O''$
    shall have no difference but in the declaration of the type that corresponds
    to $ENUM$, where we expect that $O''$ reflects the value $NEWVAL$ of $P''$ with the
    proper ordinal value, while $O'$ lacks it.
    \vspace{1pt}
}}\vspace{5pt}

\noindent With reference to our example, let $P'$ be the test program of Figure~
\ref{lst:semdsl}. Then, the test generator can satisfy the precondition of the
metamorphic relation by generating a test program $P''$ that is identical to
Figure~\ref{lst:semdsl} but in which it replaces line~\ref{lstln:semdsl:state}
with the following:
\begin{lstlisting}[
    language = dsl,
    basicstyle = \scriptsize,
    frame=single,
    numbers=none,
]
state: Enumeration { |\textbf{NEWVAL}|, Red, Green, Yellow }
\end{lstlisting}
where $NEWVAL$ is the identifier of an additional value of the enumeration. 
Next, the test generator would run the transpiler on both test programs and compute
%Our technique would then pick one of these options and construct the corresponding
%\textit{follow-up program}. Once both the \textit{source program} and the
%\textit{follow-up program} are translated into the corresponding outputs, 
the textual diff between the C code of the output sources, to crosscheck that the
postcondition stated in the metamorphic relation correctly holds. With the buggy
transpiler of our example, the textual diff would appear as follows:
\begin{lstlisting}[
    language = diff,
    basicstyle = \scriptsize,
    numbers=none,
]
@@ trafficlight.h; -7,5 +7,6 @@
 typedef enum {
   Yellow = 0,
   Green,
   Red,
+  NEWVAL,|\label{lstln:semdiff:add}|
 } trafficlight_state_t;
\end{lstlisting}
%
%gets computed and analyzed, in order to verify compliance.
%
%As an example, let us consider the third possibility of the ones outlined above
%and analyze its results. Computing the difference shows no results, since the file
%ends without changes: the \texttt{NEW} element, in fact, gets "eaten" by the compiler.
%This in turn leads to the difference being empty, which trivially satisfies the
%invariants. Nevertheless, the technique can identify the fault thanks to the lack
%of differences: namely, it is unable to find the newly added \texttt{NEW} element
%within the enumeration definition. 
At this point, the test generator would signal a failure report, noticing that the
textual diff reflects the identifier $NEWVAL$ with a different ordinal value
(\texttt{3}, as it is the fourth value of the enumeration) than the expected one
(\texttt{0}, as it was added as the first element of the enumeration).

\begin{comment}
\begin{lstlisting}[
    language = diff,
    caption = The difference between source output and follow-up output for the
              fourth case,
    label = {lst:semdiff},
    basicstyle = \scriptsize,
    float
]
@@ -6,4 +6,4 @@
 typedef enum tag_enum_state_set {
   Red = 0,
-  Green,|\label{lstln:semdiff:rem}|
+  NEW|\label{lstln:semdiff:add}|
 } state_t;
\end{lstlisting}
\end{comment}

%We will also analyze a less trivial case, represented by the fourth possibility. In
%this case, the process repeats but the difference computed is the one represented by
%Figure~\ref{lst:semdiff}. In this case, we can see the the requirements imposed by
%the \textit{differences} section of the metamorphic relation is satisfied, as shown
%by line~\ref{lstln:semdiff:add}. On the other hand, the presence of line~
%\ref{lstln:semdiff:rem} causes a violation of the \textit{invariants}, as one of the
%elements of the enumeration has been removed when it should not have. This
%situation also causes an error to be raised, signaling the failure of the oracle.

In the next section, we define our technique for metamorphic testing of transpilers in detail.

\section{The \acronym Approach to \label{sec:mtt}
Metamorphic Testing of Transpilers}
Our approach to metamorphic testing of transpilers relies on metamorphic relations that specify how \emph{mutation-style changes in the input source code} are expected to \emph{consistently map onto corresponding equivalencies and differences in the output source code} produced by the transpiler under test. We call the property that underlies these metamorphic relations \emph{mutation consistency of the (transpiled) programs} (\acronym).
In this section, we  introduce both the concept of \acronym metamorphic relations and the general workflow of a  test generator exploiting such metamorphic relations, to which we refer as \nbcc.
%Then, we discuss a set of guidelines for defining 
%\acronym metamorphic relations for traspilers.  
%, then discuss an the workflow of our automatic test generator that exploits the \acronym approach, provide a formalization on the application of metamorphic testing to transpilers. Following that, we will delineate the technique we developed as a whole and then we will delve into each stage in more detail.

\subsection{\acronym Metamorphic Relations}
\label{sec:mtt:formalize}
%For testing a given transpiler, our approach relies on metamorphic relations that specify how \emph{mutation-style changes in the input source code} are expected to correspond to \emph{consistency equivalencies and differences in the output sources} produced by the transpiler under test. We call the property that underlie these metamorphic relations \emph{mutation consistency of the (transpiled) programs} (\acronym).

An \acronym metamorphic relation can be defined as a pair
$\langle$\textit{mutation operator}, \textit{consistency checkers}$\rangle$, capturing the metamorphic relation operationally. Namely, a \emph{mutation operator} represents a punctual transformation that can be applied to an input source code, say $i$, to produce a follow-up source code, say $i'$, such that $i'$ is equal to $i$ except for the change introduced by the transformation. 
Correspondingly, the \emph{consistency checkers} specify a set of checks for validating code-level equivalences and code-level differences that we expect to
hold between the output source code that the transpiler yields when executed against the source code $i$ and $i'$, respectively,  consistently with the change done according to the given mutation operator.

\acronym metamorphic testing leverages \acronym metamorphic relations for steering the generation of test cases, while automatically assessing their outcomes through  oracles instantiated from the metamorphic relations.

%\luca{Readbility check.}
Formally, an \acronym metamorphic relation $\langle$\textit{mutation operator}, \textit{consistency checkers}$\rangle$ enables test oracles 
specifying that for each pair of input sources, $\langle i, i'\rangle$, which relate with each other as they mutually differ only for the changes determined by the given mutation operator, i.e.,  $i'=\textit{mutation operator}(i)$, then the corresponding pair of output sources produced by the transpiler $T$ under test, $\langle o=T(i),\ o'=T(i')\rangle$, must also be related with each other as they shall reflect the specific set of expected code-level equivalencies and differences, i.e.,  $\textit{consistency checkers}(o, o')$ must hold.

For instance, the following \acronym metamorphic relation  captures the metamorphic relation that we informally introduced in
Section~\ref{sec:example:our}, relatively to adding a new value in the enumeration of the program in Figure~\ref{lst:semdsl}:

%In this section, we will formalize that relation through the use of the terminology introduced earlier. For the purposes of this example, we will also assume that no name mangling occurs in the transpiler.

\fbox{\parbox[t]{.9\columnwidth}{
    \vspace{1pt}
    \begin{center}
    \textbf{\acronym metamorphic relation: \textsc{Augment Enumeration}}
    \end{center}
    \textbf{Mutation Operator}:  given an input source $i$ accepted by the transpiler under test, generate another input source $i'$ by adding a new value identifier $NEWVAL$ to an
    \texttt{Enumeration} entity that is part of an input source $i$. %, in a valid position with respect   to the ordering of the preexisting value identifiers.

    \vspace{5pt}
    \textbf{Consistency Checkers:}
    
    \begin{inparaenum}[(i)]
    \item \textit{Equivalence Check}: the output sources  $o$ and $o'$ consist of exactly equal C code, except for the \texttt{enum} definition translating the mutated \texttt{Enumeration}.
    
    \item \textit{Difference Check}: the output sources $o$ and $o'$ differ in that the \texttt{enum} definition translating the mutated \texttt{Enumeration} in $o'$ includes the value identifier $NEWVAL$, preceded (respectively, followed) by the value identifier that immediately precedes (respectively, follows)
    $NEWVAL$ in the mutated \texttt{Enumeration} in $i'$, if any.
  %  must contain a single addition in the area corresponding to the mutated \texttt{Set} matching  the following regex: \texttt{\textasciicircum~*}$NEWVAL$\texttt{,?\$}.
%    \item \textit{difference check}: the identifier $NEWVAL$ must be preceded in the diff context         by another identifier that corresponds to the attribute immediately preceding    $NEWVAL$ in the mutated \texttt{Set}, if any.
    %\item \textit{Diff check}: the identifier $NEWVAL$ must be followed in the diff context by another identifier that corresponds to the attribute immediately following       $NEWVAL$ in the mutated \texttt{Set}, if any.
    \end{inparaenum}
    \vspace{1pt}
}}%\vspace{5pt}

\subsection{\acronym Metamorphic Testing}
\begin{comment}
\begin{figure*}[tb]
\centering
\includesvg[width = \textwidth]{Nabucco Workflow.svg}
\caption{\label{fig:nbccwrkflw} Overview of the Nabucco workflow}
\Description[A graph representing the overall workflow of the Nabucco tool]{
%
%
%
% This needs doing!
%
%
%
}
\end{figure*}
\end{comment}

\begin{algorithm}[tb]
\SetKwData{SourceProgram}{$i$}
\SetKwData{FollowUpProgram}{$i'$}
\SetKwData{SourceOutput}{$o$}
\SetKwData{FollowUpOutput}{$o'$}
\SetKwData{MetamorphicRule}{$mr$}
\SetKwData{MetamorphicTransform}{\MetamorphicRule$.t$}
\SetKwData{MetamorphicCheckers}{\MetamorphicRule$.C$}
\SetKwData{MetamorphicChecker}{$c$}
\SetKwData{Transpiler}{$transpiler$}
\SetKwData{Grammar}{$g$}
\SetKwData{OracleTmp}{outcome}
\SetKwFunction{GenerateSource}{SelectInputProgram}
\SetKwFunction{MayContinue}{CanContinue}
\SetKwFunction{PickOne}{PickOne}
\SetKwFunction{Mutate}{MutateProgram}
\SetKwFunction{Invoke}{InvokeTranspiler}
\SetKwFunction{Spots}{FindSpots}
\SetKwFunction{Oracle}{VerifyOracles}
\SetKwInOut{Input}{Input}
\SetKwInOut{Output}{Output}
\Input{
    \Transpiler, the transpiler under test\\
    $MR$,  the set of \acronym metamorphic relations\\
    $S$,  a set of \inputSrc\\
    $max\_time$, maximum testing budget\\
    $max\_mr$, num.\ of $mr\in MR$ to consider, per input\\ source\\
    $max\_spots$, num.\ of applications of each $mr\in MR$
}
%\Input{
%\begin{itemize}
%    \item \Transpiler, the transpiler under test
%    \item $MR$,  the set of metamorphic relations
%    \item $S$,  a set of input sources
%    \item $max\_time$, maximum testing budget
%    \item $max\_mr$, num.\ of $mr\in MR$ to consider, per input source 
%    \item $max\_spots$, num.\ of applications of each $mr\in MR$
    %testing time, attempts across the available metamorphic relations, attempts of each metamorphic relation 
%\end{itemize}
    %a set of metamorphic relations $S=\{R = \langle t,C \rangle\}$ where $t$ is a transform operator and $C$ is a set of checkers,   a transpiler $T$
%}
\Output{$F$, the set of identified failures}
\BlankLine

$F\leftarrow\emptyset$\;
\While{\MayContinue{max\_time}}{
    \SourceProgram$\leftarrow$\PickOne{S}\;\label{alg:nbccwkrflw:isel}
%    $W\leftarrow S$\;
%    $doneMRs\leftarrow \emptyset$\;
    \While{\MayContinue{max\_time, max\_mr}}{\label{alg:nbccwkrflw:wrl}
        \MetamorphicRule$\leftarrow$\PickOne{MR}\;\label{alg:nbccwrkflw:rsel}
        $SPOTS\leftarrow$\Spots{\MetamorphicRule,\ \SourceProgram}\;\label{alg:nbccwrkflw:slst}
%        $doneSpots\leftarrow\emptyset$\;
        \While{\MayContinue{max\_time, max\_spots}}{
            $spot\leftarrow$\PickOne{SPOTS}\;\label{alg:nbccwrkflw:ssel}
            \FollowUpProgram$\leftarrow$\Mutate{\SourceProgram,\ \MetamorphicRule,\ spot}\;\label{alg:nbccwrkflw:mut}
            \SourceOutput$\leftarrow$\Invoke{\Transpiler,\ \SourceProgram}\;\label{alg:nbccwrkflw:ts}
            \FollowUpOutput$\leftarrow$\Invoke{\Transpiler,\ \FollowUpProgram}\;\label{alg:nbccwrkflw:tf}
%            \OracleTmp$\leftarrow$\Oracle{\SourceOutput,\ \FollowUpOutput, \MetamorphicRule}\;
            \If{$\neg$\Oracle{\SourceOutput,\ \FollowUpOutput,\ \MetamorphicRule}}{\label{alg:nbccwrkflw:oc}
                $F\leftarrow F\cup \langle$\SourceProgram$,\ $\FollowUpProgram$,\ $\MetamorphicRule$\rangle$\;
            }\label{alg:nbccwrkflw:ifend}
%            $doneSpots\leftarrow doneSpots \cup \{spot\}$\;
        }
%        $doneMRs \leftarrow doneMRs \cup \{MR\}$\;
    }
}
\Return{$F$}\;
\caption{\label{alg:nbccwrkflw}\nbcc's workflow}
\end{algorithm}

Algorithm~\ref{alg:nbccwrkflw} outlines the general workflow of our technique for \acronym metamorphic testing of transpilers, to which we refer to as \nbcc.
%the fuzz transpiler testing technique that we propose. We have dubbed both the technique and the corresponding research prototype \nbcc.
%In particular, the prototype performs fuzz testing of a transpiler in accordance with our metamorphic testing approach.

\nbcc takes as input a transpiler under test, a set of \acronym metamorphic relations valid for the given transpiler, a set of \inputSrc accepted by the transpiler, and three hyper-parameters that control the testing budget allowance and its segmentation across the given metamorphic relations and seed inputs. 
\nbcc proceeds through 
%can be conceptualized as 
three nested loops, which iterate through the \inputSrc, the \acronym metamorphic relations and the possible ways of applying mutations, respectively, as follows. 

%$S$ which will be used as the base code to which our metamorphic approach applies the mutations. Formally, the cardinality of this set can be either finite or infinite, with the former representing a set of programs that have been pre-generated prior to the execution of our technique, and the latter modeling an on-demand generation approach according to a desired strategy. 

%Additionally, the technique requires a set of \acronym  metamorphic relations to be applied $MR$, according to the formalization described in Section~\ref{sec:mtt:formalize}. Finally, $max\_time$, $max\_mr$ and $max\_spots$ represent budget parameters that allow to define respectively the amount of time our technique should dedicate to the testing activity, the maximum amount of \acronym metamorphic relations that need to be considered for every input source chosen, and the number of spots where the mutation described by the chosen relation may be applied to at most.

%\giovanni{Describe the inputs first}

%
In the first loop, lines 2-\ref{alg:nbccwkrflw:isel}, if further testing time is still available (parameter \textit{max\_time}), it selects one of the available \inputSrc to play the role of \baseInput (the input source $i$ following the terminology of  Section~\ref{sec:mtt:formalize}) throughout the metamorphic testing steps thereon. 
%each of which iterates until its corresponding budget is exhausted. In particular, the first loop will keep operating until the time budget is reached, in accordance with general fuzz testing approaches. During the loop, \nbcc will choose one among the input sources provided as input, as shown on line~\ref{alg:nbccwkrflw:isel}. In accordance to the terminology outlined in Section~\ref{sec:mtt}, we call the chosen input source $i$ \emph{source program}.

\begin{comment}
\nbcc~can be conceptualized as three nested loops, each of which iterates on a specific
component\enea{? better words are welcome} until its corresponding budget is exhausted. In
particular, the first loop of the approach will keep operating until a global time budget
limit provided as an input by the user is exhausted. During the loop, \nbcc~will choose
one among the input sources provided as input, as shown on line~\ref{alg:nbccwkrflw:isel}.
%Formally, the cardinality of the input set can be either finite or infinite, with the former
%representing a set of programs that have been pre-generated prior to the execution of the
%technique and the latter modeling an on-demand generation approach according to a desired
%strategy.
In accordance with the terminology outlined in Section~\ref{sec:mtt}, we call
the chosen input source $i$ \emph{source program}.
\end{comment}

In the second loop, lines~\ref{alg:nbccwkrflw:wrl}-6, if there is further testing time and up to a maximum number of attempts (parameter \textit{max\_mr}), 
\nbcc selects an \acronym metamorphic relation (line~\ref{alg:nbccwkrflw:wrl}) to steer the test generation process. It then inspects the \baseInput $i$ (line~6), searching for all possible locations where the mutation operator of the metamorphic relation can be actually applied. The found locations are called \emph{spots} in the algorithm.
%provided as an input, picking one metamorphic relation $mr$ out of the available ones according to a certain strategy. And then, \nabcc inspects the relation's \emph{mutation operator} and the \emph{source program} under consideration to identify all the locations where the specified mutation can be applied. We call these locations \emph{spots}.

For instance, if \nbcc is testing the sample (bugged)
transpiler of our working example, assuming that in the first loop it selected the  source program of Figure~\ref{lst:semdsl} as seed input source, and assuming that in the second loop it selected the \acronym metamorphic relation exemplified in Section~\ref{sec:mtt:formalize},   
%and the single source program of Figure~\ref{lst:semdsl}.
then it will determine that, according to the mutation operator of the metamorphic relation at hand, the new
value identifier $NEWVAL$ can be added at four possible spots ($\S$ the set $SPOTS$ at line 6 of Algorithm~\ref{alg:nbccwrkflw}): 
\begin{inparaenum}[(i)]
\item before the \texttt{Red} identifier, \label{lst:wkflwx:bred}
\item between the \texttt{Red} and \texttt{Green} identifiers,
\item between the \texttt{Green} and \texttt{Blue} identifiers,
or \item after the \texttt{Blue} identifier.
\end{inparaenum}

Eventually, in the third loop, lines 7-14, if there is further testing time and up to a maximum number of attempts (parameter \emph{max\_spots}), \nbcc executes the actual metamorphic testing steps. It selects a possible spot (line~\ref{alg:nbccwrkflw:ssel}), generates a follow up input source $i'$ by applying the mutation operator of the metamorphic relation at hand at the given spot (line~\ref{alg:nbccwrkflw:mut}), executes the transpiler against both the input sources $i$ and $i'$ (lines~\ref{alg:nbccwrkflw:ts} and~\ref{alg:nbccwrkflw:tf}, respectively) and verifies the test oracles against the outcomes from the transpiler, $o$ and $o'$,  according to the consistency checkers of the metamorphic relation at hand (line~\ref{alg:nbccwrkflw:oc}). 
%iterates on the identified spots. The function call on line~\ref{alg:nbccwrkflw:ssel} represents the strategy used to select a spot of interest among the available ones. 
%The \emph{source program} is then mutated according to the \emph{mutation operator} of $mr$, injecting the modifications in the chosen spot to obtain the \emph{follow-up program} $i'$ (line~\ref{alg:nbccwrkflw:mut}). Both the source
%and the follow-up programs are then translated through invocation of the transpiler, as shown on lines~\ref{alg:nbccwrkflw:ts} and~\ref{alg:nbccwrkflw:tf}, to obtain the corresponding outputs, i.e. the \emph{source output} and the \emph{follow-up output}.
%Finally, \nbcc verifies whether the oracles represented by $mr$'s consistency checkers are verified or not through comparison of the source and follow-up outputs
%(line~\ref{alg:nbccwrkflw:oc}). 
Finally, it collects all failing test cases (line~13) and reports them as output at the end (line~18).
%If a failure is detected, then the metamorphic relation under test, along with both the \emph{source program} and the \emph{follow-up program} are tracked to be reported later.
%On exhaustion of the available time budget, the fuzzing activity carried out by \nbcc terminates, and the failures are reported back to the user for investigation.

Continuing with the example, \nabcc iteratively applies the metamorphic testing steps by inserting the new value identifier \texttt{NEWVAL} at the various possible spots of the seed input source of Figure~\ref{lst:semdsl}.

%Continuing with the example, \nabcc would iterate metamorphic testing steps by adding the new value identifier \texttt{NEWVAL} at the possible spots of the source program of Figure~\ref{lst:semdsl}.

Eventually, it might 
%$pick the spot~\ref{lst:wkflwx:bred}, adding 
add the identifier $NEWVAL$ 
as the first value of the \texttt{Enumeration} in the program, 
%on line~\ref{lstln:semdsl:state} of Figure~\ref{lst:semdsl}. 
and observe the failure of the test oracles, %as 
%The consistence checker would then identify a failure, in particular during the execution of the relation's \emph{difference check}, 
as the
value \texttt{Red} that immediately follows $NEWVAL$ in the follow-up input source $i'$ does not appear
in the expected location in the transpiled output $o'$. 
Consequently, \nbcc would report a test failure.% accordingly.

Algorithm~\ref{alg:nbccwrkflw} describes the general workflow of \nbcc, but several extensions and optimizations could be easily incorporated.
%straight away.
%
The set of \inputSrc could comprise existing input sources, but also input sources automatically generated, or a mix of existing and automatically generated input sources. It may even be incrementally augmented with the follow-up input sources generated at line~\ref{alg:nbccwrkflw:mut}, or with a selection of these follow-up input sources, chosen according to strategies aimed at identifying those more interesting for further mutations in subsequent testing steps.
%
%possible strategies for identifying which follow-up input sources are interesting to be considered for further mutations at next testing steps. 
%
The strategies for iterating through the \inputSrc, the metamorphic relations and the mutation spots (the \texttt{PickOne} functions in the algorithm) can also be customized, for example to avoid picking the same items multiple times.
Finally, the translation of the source program (line~\ref{alg:nbccwrkflw:ts}) could also be made more efficient. In fact, it is entirely independent of the metamorphic relation and picked spot, and can therefore be executed immediately after the choice of the input source (line~\ref{alg:nbccwkrflw:isel}), while caching its result.
%Yet another form of optimization involves the translation of the source program on line~\ref{alg:nbccwrkflw:ts}, which is entirely independent of the metamorphic relation and picked spot. We can therefore cache the result of the transpilation by executing it immediately after the choice of the input source (line~\ref{alg:nbccwkrflw:isel}).

Our current implementation\footnote{\url{https://doi.org/10.5281/zenodo.19340835}.} works with automatically generated \inputSrc, and  iterates through \inputSrc, metamorphic relations and mutation spots in random order, caching the corresponding selections to avoid re-picking, and resetting each cache upon exhausting all available options at a given selection point.
Additionally our implementation already implements the optimization involving the early translation of the source program and its caching.
We leave for future work to investigate possible strategies for feeding the follow-up input sources as additional seeded inputs.

\section{Case Study and Empirical Evaluation}
\label{sec:casestudy}
%\luca{Introdurre al contesto del partner industriale. Spiegando un solo esempio e descrivendo gli altri ad alto livello.}
We devised and investigated the \acronym approach in the context of a technology-transfer project joint with an industrial partner. As we already mentioned in Section~\ref{sec:example}, the industrial partner worked with a 
%In our work, we considered as a case study 
custom DSL for defining control logics of industrial plants, by using a state-machine-based formalism and according to a syntax that is pervasively inspired by the natural-language terminology of their engineers, in the fashion of the sample program of Figure~\ref{lst:semdsl}. The industrial partner used a transpiler developed in Java in their laboratories to translate the DSL control logics into C programs.   
%created by an industrial partner for the management of railway networks. Similarly to our example described
%in Section~\ref{sec:example}, 
%the language makes use of a syntax that is heavily influenced by natural language constructs. As mentioned, the input language describes plant \enea{layout?} logic by means of a finite state machine,  enumerating the states the plant can assume and the transitions between them. 

In this section we describe a case study, in which we specialized the \acronym metamorphic testing approach to the transpiler developed by our industrial partner. For confidentiality reasons, we cannot disclose  the identity of the partner, nor the technical details of their transpiler or their DSL. Thus, we present the \acronym metamorphic relations discussed in the remainder of this section by referring to an anonymized version of the partner’s DSL, mimicking the DSL that we already introduced in the working example in Figure~\ref{lst:semdsl}, while nonetheless keeping strict correspondence with the characteristics of the actual constructs of the partner’s DSL and the way we addressed those constructs in the actual setting of the case study.

\subsection{Instantiating \acronym Metamorphic Relations}

\begin{figure*}[h!]
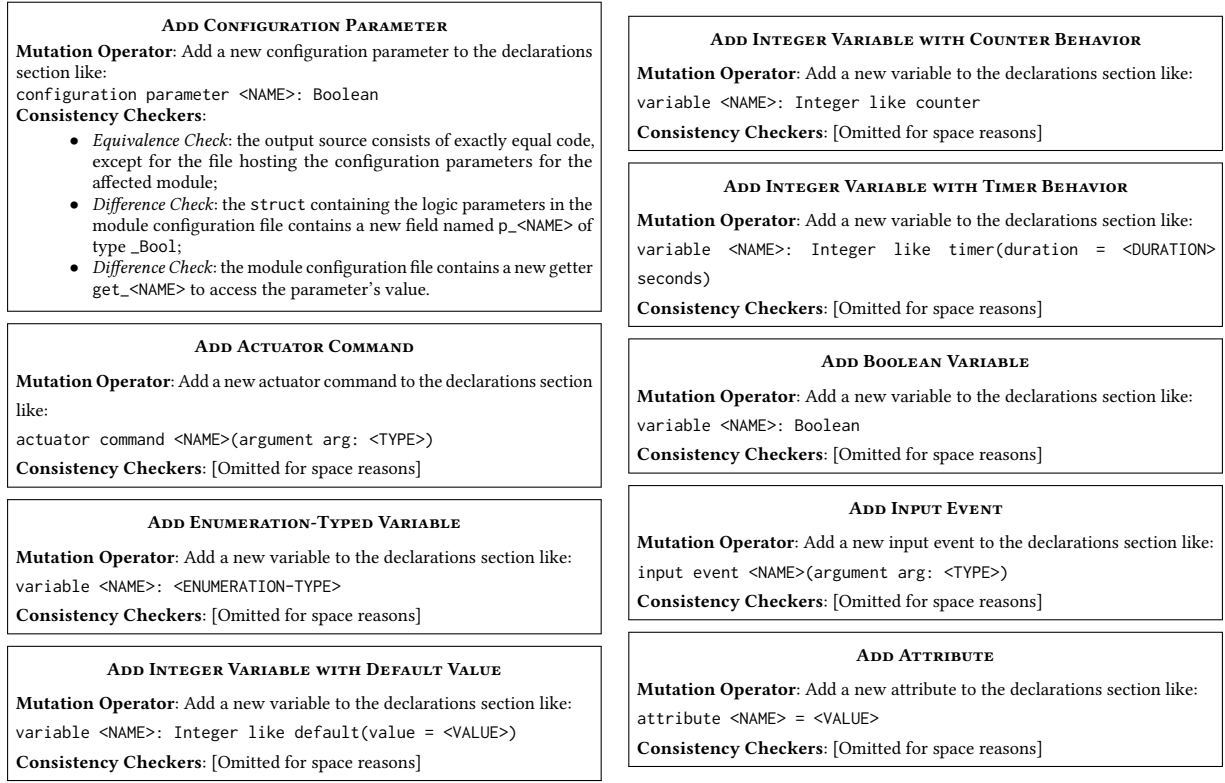

\begin{center}
\begin{tabular}{c c}
\makecell[c]{
\fbox{\parbox[t]{.9\columnwidth}{
\begin{footnotesize}
    \vspace{1pt}
    \begin{center}
    \textbf{\confParamMRName}
    \end{center}
    \textbf{Mutation Operator}:
    Add a new configuration parameter to the declarations section like:\\
    \texttt{configuration parameter <NAME>: Boolean}\\
    \textbf{Consistency Checkers}:
    \begin{itemize}
    \item \textit{Equivalence Check}: the output source consists of exactly equal code, except
          for the file hosting the configuration parameters for the affected module;
    \item \textit{Difference Check}: the \texttt{struct} containing the logic parameters in the
          module configuration file contains a new field named \texttt{p\_<NAME>} of type
          \texttt{\_Bool};
    \item \textit{Difference Check}: the module configuration file contains a new
          getter \texttt{get\_<NAME>} to access the parameter's value.
    \end{itemize}
\end{footnotesize}
}}\\
\fbox{\parbox[t]{.9\columnwidth}{
\begin{footnotesize}
    \vspace{1pt}
    \begin{center}
    \textbf{\textsc{Add Actuator Command}}
    \end{center}
    \textbf{Mutation Operator}:
    Add a new actuator command to the declarations section like:\\
    \texttt{actuator command <NAME>(argument arg: <TYPE>)}\\
    \textbf{Consistency Checkers}: [Omitted for space reasons]
\end{footnotesize}
}}\\
\fbox{\parbox[t]{.9\columnwidth}{
\begin{footnotesize}
    \vspace{1pt}
    \begin{center}
    \textbf{\textsc{Add Enumeration-Typed Variable}}
    \end{center}
    \textbf{Mutation Operator}:
    Add a new variable to the declarations section like:\\
    \texttt{variable <NAME>: <ENUMERATION-TYPE>}\\
    \textbf{Consistency Checkers}: [Omitted for space reasons]
\end{footnotesize}
}}\\
\fbox{\parbox[t]{.9\columnwidth}{
\begin{footnotesize}
    \vspace{1pt}
    \begin{center}
    \textbf{\textsc{Add Integer Variable with Default Value}}
    \end{center}
    \textbf{Mutation Operator}:
    Add a new variable to the declarations section like:\\
    \texttt{variable <NAME>: Integer like default(value = <VALUE>)}\\
    \textbf{Consistency Checkers}: [Omitted for space reasons]
\end{footnotesize}
}}}

&

\makecell[c]{
\fbox{\parbox[t]{.9\columnwidth}{
\begin{footnotesize}
    \vspace{1pt}
    \begin{center}
    \textbf{\textsc{Add Integer Variable with Counter Behavior}}
    \end{center}
    \textbf{Mutation Operator}:
    Add a new variable to the declarations section like:\\
    \texttt{variable <NAME>: Integer like counter}\\
    \textbf{Consistency Checkers}: [Omitted for space reasons]
\end{footnotesize}
}}\\
\fbox{\parbox[t]{.9\columnwidth}{
\begin{footnotesize}
    \vspace{1pt}
    \begin{center}
    \textbf{\textsc{Add Integer Variable with Timer Behavior}}
    \end{center}
    \textbf{Mutation Operator}:
    Add a new variable to the declarations section like:\\
    \texttt{variable <NAME>: Integer like timer(duration = <DURATION> seconds)}\\
    \textbf{Consistency Checkers}: [Omitted for space reasons]
    %\begin{itemize}
    %\item \textit{Equivalence Check}: the output sources consist of exactly equal code, except
    %      for the files hosting the \texttt{struct} and the function implementations of the
    %      affected logic;
    %\item \textit{Difference Check}: the \texttt{struct} containing the logic variables in
    %      the logic's source files contains a new field named \texttt{<NAME>} of type
    %      \texttt{timer\_t};
    %\item \textit{Difference Check}: the logic source files contain a getter
    %      \texttt{Get<NAME>} to access the timer's current value;
    %\item \textit{Difference Check}: the logic source files contain a call to
    %      \texttt{timer\_init} in their initialization logic, passing \texttt{<NAME>} and
    %      \texttt{<DURATION>} as parameters;
    %\item \textit{Difference Check}: the logic source files contain a call to 
    %      \texttt{timer\_exec} in their tick logic, passing \texttt{<NAME>} as parameter.
    %\end{itemize}
\end{footnotesize}
}}\\
\fbox{\parbox[t]{.9\columnwidth}{
\begin{footnotesize}
    \vspace{1pt}
    \begin{center}
    \textbf{\textsc{Add Boolean Variable}}
    \end{center}
    \textbf{Mutation Operator}:
    Add a new variable to the declarations section like:\\
    \texttt{variable <NAME>: Boolean}\\
    \textbf{Consistency Checkers}: [Omitted for space reasons]
\end{footnotesize}
}}\\
\fbox{\parbox[t]{.9\columnwidth}{
\begin{footnotesize}
    \vspace{1pt}
    \begin{center}
    \textbf{\textsc{Add Input Event}}
    \end{center}
    \textbf{Mutation Operator}: 
    Add a new input event to the declarations section like:\\
    \texttt{input event <NAME>(argument arg: <TYPE>)}\\
    \textbf{Consistency Checkers}: [Omitted for space reasons]
\end{footnotesize}
}}\\
\fbox{\parbox[t]{.9\columnwidth}{
\begin{footnotesize}
    \vspace{1pt}
    \begin{center}
    \textbf{\textsc{Add Attribute}}
    \end{center}
    \textbf{Mutation Operator}:
    Add a new attribute to the declarations section like:\\
    \texttt{attribute <NAME> = <VALUE>}\\
    \textbf{Consistency Checkers}: [Omitted for space reasons]
\end{footnotesize}
}}\\
}
\end{tabular}
\end{center}
\caption{The \acronym metamorphic relations instantiated in our case study}
\label{fig:cs:mr} 
\end{figure*}

In the case study we instantiated a set of \acronym metamorphic relations for the DSL constructs
that can appear in the \texttt{declarations} section of a control-logic definition. A subset of the constructs are 
exemplified in the \texttt{declarations} section of the sample program in
Figure~\ref{lst:semdsl},
% An example of how this section can be modeled is shown in.
%In fact, this section is also illustrated in the example of Figure~\ref{lst:semdsl}, which includes sample definitions for several of the constructs for which we defined corresponding \acronym metamorphic relations.
while in Figure~\ref{fig:cs:mr} we report the nine \acronym metamorphic relations that we derived for the possible constructs, referring to the DSL grammar and the documented requirements of the transpiler.

As explained in Section~\ref{sec:mtt:formalize}, each metamorphic relation consists of a \emph{mutation operator} and a set of \emph{consistency checkers}. In the figure, we documented the mutation operator for all nine metamorphic relations, while, due to the space limitations, we documented the consistency checkers only for the first metamorphic relation. The omitted consistency checkers are available for reference in the repository of our tool.\footnote{\url{https://doi.org/10.5281/zenodo.19340835}.}  

In the figure, the titles of the metamorphic relations refer to the behavior of the corresponding mutation operators, each handling the injection of a construct that can be part of the \texttt{declarations} section of a control-logic definition. 
The first three metamorphic relations (\emph{Add configuration parameter}, \emph{Add actuator command} and \emph{Add enumeration-typed variable}) correspond to the constructs actually exemplified in the sample program in
Figure~\ref{lst:semdsl}, whereas the other matamorphic relations refer to other types of variables or other types of entities (namely, input events and attributes) that can be also added. 

The mutation operators are documented in the first paragraph of the description of the metamorphic relations, 
%is composed by a
%description of the possible injection spots, expressed through the construct name and the code region as described in the DSL grammar, and a 
including a template code snippets representing the grammatical structure of the construct to be added.
For instance, for the first metamorphic relation, \emph{Add configuration parameter}, 
the mutation operator augments the \texttt{declarations} section by adding (at a valid position of the section) a line that starts with the keyword 
%in the top-left of the figure. The description indicates that the construct to add is called
\emph{configuration parameter}, followed by a valid name (\emph{<NAME>}) of the parameter, which must be of type \emph{Boolean}. This produce a declaration that respects the same syntax as the one at Figure~\ref{lst:semdsl}, line~\ref{lstln:semdsl:camparam}.
%The valid spots for applying the mutation operator are at any possible position in the the end of every line in the \texttt{declarations} block.
%of our DSL ($\S$ Figure~\ref{lst:semdsl}, lines~\ref{lstln:semdsl:decbegin}\textendash\ref{lstln:semdsl:camparam}).
%Analysis of the DSL's grammar shows that the targeted construct can be added in-between any of the other constructs already present in the \texttt{declarations} block of our DSL ($\S$ Figure~\ref{lst:semdsl}, lines~\ref{lstln:semdsl:decbegin}\textendash\ref{lstln:semdsl:decend}). This description identifies as valid spots the end of every line from~\ref{lstln:semdsl:decbegin} to~\ref{lstln:semdsl:camparam}.
%The template code snippet specifies the exact syntax to use to add the new configuration parameter. The template contains \texttt{<NAME>} as a placeholder, which will be replaced by a valid identifier during instantiation of the metamorphic relation, say \texttt{MCPTEST\_a26c5}. 
%The snippet to add to the code would thus become \texttt{configuration parameter MCPTEST\_a26c5: Boolean}, matching the already existing construct in syntax ($\S$ Figure~\ref{lst:semdsl}, line~\ref{lstln:semdsl:camparam}).

The second paragraph of the metamorphic relation \emph{Add configuration parameter} exemplifies the documention of 
%We then proceed with the definition of 
the \emph{consistency checkers}.  Each checker can represent either an  \emph{equivalence check} or a \emph{difference check}. The formers verify that given code areas are unchanged across the C sources yielded by the transpiler for the base and follow-up test cases, respectively, while the latter ones predicate on the differences found therein. 
%We call these, respectively, \emph{equivalence checks} and \emph{difference checks}. 
For the metamorphic relation \emph{Add configuration parameter}, we specified three checkers, documenting the expected equivalences and differences, with reference to the C code that the considered transpiler produces as output.  
%Considering once again the example of \confParamMRName, we identified three checkers, 
%which we exemplify by referencing Figure~\ref{lst:semcbug} and the previously mentioned instantiation of the mutation operator. 
The first consistency checker specify an equivalence check ensuring that the addition of a configuration parameter impacts only on the relevant files.
%not impact any file except for the ones shown in the Figure~\ref{fig}.
This enables the detection of failures arising from faults in the transpiler for which the specification of a configuration parameter yields unwanted cascading effects on the other files of the translated source code.
%This allows us to expose failures if the transpiler results has a bug such that the specification of a configuration parameter  
%that do not impact the translation of the affected construct but 
%yields unwanted cascading effects on the other files of the translated source code.

The second consistency checker verifies that the boolean parameter has been properly added to the \texttt{struct} in the proper header file.
%e.g., the header file $\S$ \texttt{trafficlight.h}, line~\ref{lstln:semcbug:parstructbegin}), with the appropriate boolean type and the name \texttt{p\_MCPTEST\_a26c5} as determined during the instantiation. 
Similarly, the third consistency checker makes sure that the transpiler has added a proper getter in the produced C code.
%both files named after the parameter (\texttt{get\_MCPTEST\_a26c5}) and with the proper code structure ($\S$ \texttt{trafficlight.h}, line~\ref{lstln:semcbug:parget} and \texttt{trafficlight.c}, line~\ref{lstln:semcbug:pargetc}). 
For instance, if the mutation operator of the metamorphic relation \emph{Add configuration parameter} is applied to add a line as:
\begin{lstlisting}[
    language = dsl,
    basicstyle = \footnotesize,
    numbers=none,
]
configuration parameter MCPTEST_a26c5: Boolean
\end{lstlisting}
at the end of the \texttt{declarations} section in the sample code of Figure~\ref{lst:semdsl}, then the difference checks verify that the 
diff-snippet between the C sources yielded by the transpiler for the base and follow-up test case, respectively, matches the following differences:
\begin{lstlisting}[
    language = diff,
    basicstyle = \scriptsize,
    numbers=none,
]
@@ trafficlight.h; -4,3 +4,4 @@
 typedef struct {
   _Bool p_enableCamera;
+  _Bool p_MCPTEST_a26c5;
 } trafficlight_t;
@@ trafficlight.h; -14,0 +14,1 @@
+ _Bool get_MCPTEST_a26c5(trafficlight_t *);
@@ trafficlight.c; -23,0 +23,3 @@
+ _Bool get_MCPTEST_a26c5(trafficlight_t *instance) {
+   return instance->p_MCPTEST_a26c5;
+ }
\end{lstlisting}

\subsection{Research Questions}
\label{sec:exp}
In the case study, we  executed \nbcc against the partner's transpiler, relying on the  metamorphic relations defined above. Our investigation was driven by two main research questions:

\begin{enumerate}[label=\textbf{RQ\arabic*}:]
\item Can \nbcc effectively detect faults?
\item Can \nbcc outperform a baseline that only fuzzes the transpiler?

%er that exploits the mutations of our \acronym metamorphic relations?
\end{enumerate}

Question \textbf{RQ1} studies the effectiveness of 
our  metamorphic testing approach for transpilers, which relies on the test cases and the test oracles induced by the \acronym metamorphic relations.
We addressed this research question by seeding faults into the transpiler and measuring how many of those faults \nbcc can  reveal, as well as the portion of detected faults that were specifically exposed thanks to the metamorphic oracles.

%corresponds to a study in which we inject potential faults into the software and verify if they can be revealed thanks to our metamorphic oracles. This process requires an initial injection of the faults and verification only for the follow-up program, then a confirmation step is performed to verify that these faults can be revealed if applied to the source program, too.

Question \textbf{RQ2} corresponds to an ablation study in which we adapt \nbcc to work as a fuzzer that applies the same mutations used in \nbcc, but without exploiting knowledge of the underlying metamorphic relations and oracles. 
This eliminates the overhead that \nbcc pays for checking the metamorphic oracles, allowing the generator to spend the saved time on additional fuzzing.
In this way, we assess whether the overhead introduced by \acronym metamorphic testing is justified by the benefits it provides over pure fuzzing.

%For \textbf{RQ2}, we decided to compare \nbcc~against
%the baseline testing technique of 
%random testing, as we have not identified any other testing technique that could be applied to the specifics of transpilers in a way similar to ours.

\subsection{Experimental Settings}

We developed \nbcc in Java. It provides an API to customize the \acronym metamorphic relations accessible both from Java and through YAML configuration files. The transpiler under test is specified in a YAML configuration file. 
The tool is open source and accessible anonymously for reviewers.\footnote{\url{https://doi.org/10.5281/zenodo.19340835}.}

For the case study, we connected \nbcc to the transpiler under test, and configured it to handle the nine metamorphic relations of Figure~\ref{fig:cs:mr}. To increase the throughput of test cases  per metamorphic relation, we exploited parallelism by distributing the nine relations across four parallel instances of the tool. 
Each instance was configured to handle a mutually disjoint subset of metamorphic relations (3, 2, 2 and 2 metamorphic relations, respectively). Accordingly, we set the \texttt{max\_mr} parameter of each instance to 3, 2, 2, and 2, respectively, to ensure that every metamorphic relation is picked exactly once after selecting a seed input ($\S$Algorithm~\ref{alg:nbccwrkflw}, line~\ref{alg:nbccwkrflw:wrl}).
Furthermore, we set the \texttt{max\_spots} parameter to 4, such that each metamorphic relation is applied at four different locations within each seed input ($\S$Algorithm~\ref{alg:nbccwrkflw}, line~\ref{alg:nbccwkrflw:isel}).

We provided \nbcc with automatically generated seed inputs, relying on the tool rmutt.js~\cite{tachuris_tachurisrmuttjs_2024} to generate control logic sources based on the grammar of the partner’s DSL. We executed rmutt.js for 6 hours, retaining only the sources that the transpiler accepted as valid control logics. We ended up with 297 valid sources. 

A testing session consisted in  executing the four parallel instances of \nbcc configured as above  against the partner's transpiler with a time budget (\texttt{max\_time}) of 12 hours.
To account for the random choices that underlie the \nbcc algorithm, we executed \numR repetitions of the 12 hour testing session, configured as above.
During each 12‑hour testing session, the seed inputs were first randomly shuffled and then processed in the same order by all four parallel instances of \nbcc.
Each testing session has been executed on a cloud resource running Windows Server 2025 Standard 24H2 build 26100.6584, equipped with 12 processors, 32 GB of RAM and a 100 GB SSD disk.

%To verify the efficacy of the proposed technique, we implemented a prototype tool, also called \nbcc and ran it against a case study. 

%In order to parallelize our approach as much as possible while also taking into account space and concurrency concerns, we then set up four instances of our prototype \nbcc. Each instance would then be responsible for running tests on a predefined set of metamorphic relations: we manually split the previously mentioned relations across the four instances according to a 3/2/2/2 distribution. We then provided every instance with a time budget of 12 hour and a shared seed set at the beginning of each execution.

%This setup allowed us to simulate the execution of a single
%instance of \nbcc~on the target transpiler, covering all the metamorphic rules we
%had identified during a four-day testing campaign.

%\enea{Our experimental setup seems slightly similar (at least in passing) to the one by Le et al. Do we want to mention that?} Giovanni: No :)

%\subsubsection{Setup for RQ1}
%\label{sec:exp:rq1set}
%\luca{Riscrivere il seguente. Per prima cosa indicare l'obiettivo: 1) iniettare dei fault sulle condizione per validare la tecnica 2) scegliendo quelli che evitano crash troppo ovvi. 3) Al fine di identificare quelli troppo ovvi, abbiamo iniettato tutti i possibili fault usando un programma semplice (LNC vuoto), ipotizzando che i fault che fanno crashare il transpiler su quel progetto sono ovvi.}

To answer RQ1, we evaluated the effectiveness of \nbcc in detecting faults by systematically injecting seeded faults into the transpiler’s bytecode. To this end, we developed a custom bytecode instrumentor  that 
targets the opcodes implementing conditional branches (\texttt{ifeq}, \texttt{ifne}, \texttt{ifnonnull}, \texttt{ifnull}) in the core code of the transpiler (i.e.,  excluding the external libraries like ANTLR). Each injected fault forces the branch condition to evaluate deterministically to either \texttt{true} or  \texttt{false}.

While this strategy enables the generation of a large number of seeded faults, it also suffers of inefficiencies that may negatively impact the evaluation of \nbcc. On one hand, many injected faults may lead to trivial crashes of the transpiler. Such trivial crashes are both scarcely representative of realistic faults that a developer would introduce and can be straightforwardly  revealed even without any specific test oracle. On the other hand, executing every generated test case against every faulty version of the transpiler would be prohibitively expensive, also because of the large amount of test cases that \nbcc generates in the 12 hour testing sessions.

To mitigate such issues, we heuristically refined the fault injection strategy as follows.
First, in a preparatory experiment, we
executed the transpiler against a sample DSL control logic  (provided by our partner), considering every injected fault in a separate execution.  
We discarded all injected faults that caused the transpiler to crash in the preparatory experiment.

Moreover, we exploited a heuristic fault‑selection strategy that does not impact the number of test executions done by \nbcc, while maximizing the number of injected faults evaluated at least once in a testing session: It aims at maintaining that, in a testing session, we evaluate each test case against a distinct injected fault.
In detail, the intrumentor incorporates all injected faults at once (but the ones discarded in the preparatory experiments). At runtime, the instrumentation activates only the
first executable injected fault that has not been executed in any previous test cases (if any). An injected fault is executable when the execution traverses the corresponding branch.
This approach ensures broad consideration of potentially interesting faults while controlling the number of tested faulty  variants of the transpiler.

\begin{comment}
\begin{enumerate}[label=(\alph*)]
\item \textbf{The condition has not been instrumented before}: the instrumentation tool
      randomly picks between instrumenting the condition with "true" or with "false" and
      modifies the code accordingly;
\item \textbf{The condition has already been instrumented once}: the instrumentation
      tool modifies the code to do the opposite of what had been done before (if the
      condition had been instrumented with "true" it now gets instrumented with "false",
      and vice versa);
\item \textbf{The condition has already been instrumented for both "true" and "false" cases}:
      the instrumentation tool does not modify the target code and selects the next available
      condition to instrument.
\end{enumerate}
\end{comment}

%\enea{Come facciamo questa cosa? Noi lavoriamo sulle relazioni metamorfiche; facciamo
%injection solamente su follow-up; però non è detto che i fault siano replicabili anche
%nel sorgente; farlo fin da subito è costoso...}

%\luca{Spiegare che la conferma è una cosa derivata dal processo, e nella realtà non servirebbe il follow-up in quanto il bug sarebbe già presente nel source (riga alla fine di conseguenza questa verifica non fa parte del budget reale della tecnica).}

\nbcc evaluates the injected faults only while executing follow‑up test cases, recording which injected faults crashed the transpiler or
were detected by the metamorphic oracles. 
Whereas the crashes indicate faults that \nbcc would necessarily detect, failed oracles may be spurious alarms, as the base test case considered the non-faulty transpiler. 
Thus, 
%We did not track crashes at this stage, since any crash—whether triggered by the source or follow‑up program—would necessarily be detected by our approach.
%Finally, 
after completing each testing session, we performed a confirmation step: for each injected fault that \nbcc detected via a metamorphic oracle, we re‑executed the pair of test cases that revealed the fault, to confirm if the fault gets indeed revealed according to the metamorphic relations comprised by those test cases. We removed the unconfirmed outcomes. 
We remark tat this confirmation step is specific to our experimental protocol and is not part of the \nbcc technique itself; therefore, we execute the confirmation step offline with respect to the 12‑hour time budget used in the main experiments.

For RQ2, we turned \nbcc into a corresponding fuzzer version, which exploits the same mutation operators and iteratively executes  mutated test cases, although ignoring the underlying metamorphic relations and oracles. As with common fuzzers, there are no oracles, and it reports failures upon observing the transpiler crashes. We used the same configuration of the 12-hour testing sessions as above, seeding the same injected faults with the same heuristic strategy. Moreover, for each testing session executed with \nbcc, we recorded the seed of the random number generator used in the tool, and executed a corresponding fuzzing session which uses exactly the same seed. In this way, the fuzzing session shuffles the input sources in the same order as the corresponding \nbcc session, as well as it applies the mutation operators in the same order and at the same code locations. This improves the comparability of the results.

\subsection{Results}
\label{sec:exp:res}

\IgnoreSpacesOn
\prgNewFunction \recolorResultsCellColumns {} {
    \intStepOneInline{2}{\arabic{rowcount}}{
        \intIfEvenT{\intMathDivTruncate{##1}{2}}{
            \rowSetStyle{##1}{bg=lightgray!30}
        }
    }
}
\IgnoreSpacesOff

\begin{figure*}[htbp]
\begin{center}
\begin{tblr}{
    vline{3} = {1-Z}{solid},
    hline{2} = {1-Z}{solid},
    column{3-Z} = {r},
    process=\recolorResultsCellColumns
}
  &  & Transpiler runs & Injected faults & Crashes & Oracle failures \\
  \SetCell[r=2]{c} Testing session  1 & \nbcc & 1952 & 689 & 13 & 33 \\
                          & \textsc{Fuzzer} & 1952 & 692 & 13 & - \\
  \SetCell[r=2]{c} Testing session  2 & \nbcc & 2116 & 714 & 15 & 40 \\
                          & \textsc{Fuzzer} & 2128 & 714 & 13 & - \\
  \SetCell[r=2]{c} Testing session  3 & \nbcc & 352 & 134 & 5 & 6 \\
                          & \textsc{Fuzzer} & 360 & 133 & 5 & - \\
  \SetCell[r=2]{c} Testing session  4 & \nbcc & 864 & 324 & 5 & 20 \\
                          & \textsc{Fuzzer} & 864 & 324 & 5 & - \\
  \SetCell[r=2]{c} Testing session  5 & \nbcc & 2206 & 707 & 12 & 28 \\
                          & \textsc{Fuzzer} & 2194 & 707 & 12 & - \\
  \SetCell[r=2]{c} Testing session  6 & \nbcc & 2074 & 714 & 16 & 40 \\
                          & \textsc{Fuzzer} & 2086 & 714 & 14 & - \\
  \SetCell[r=2]{c} Testing session  7 & \nbcc & 2128 & 714 & 12 & 39 \\
                          & \textsc{Fuzzer} & 2120 & 714 & 12 & - \\
  \SetCell[r=2]{c} Testing session  8 & \nbcc & 2110 & 714 & 14 & 40 \\
                          & \textsc{Fuzzer} & 2110 & 714 & 14 & - \\
  \SetCell[r=2]{c} Testing session  9 & \nbcc & 1816 & 688 & 12 & 27 \\
                          & \textsc{Fuzzer} & 1816 & 693 & 13 & - \\
  \SetCell[r=2]{c} Testing session 10 & \nbcc & 2128 & 714 & 14 & 43 \\
                          & \textsc{Fuzzer} & 2128 & 714 & 14 & -
\end{tblr}
\end{center}
\caption{Results of \nbcc and the \textsc{Fuzzer} version per testing session}
\label{tab:exp:res}
\end{figure*}

Figure~\ref{tab:exp:res} reports the results of our experiments, for the \numR testing sessions of 12 hours executed with both \nbcc and the fuzzer version of the tool. 
%In particular, we performed a total of \numR runs of both \nbcc and its ablated version. 
Each pair of rows
in the table reports on a testing session, siding (first and second row or each pair or rows) the results
from the execution (with same seed of the random number generator)
%of the corresponding testing sessions (same seed of the random number generator) executed 
with either \nbcc or the fuzzer version, respectively. 
%. For each run, the top row represents the results obtained through the execution of our prototype, while the bottom row represents the data obtained from the ablation study.  For further clarification, the \textit{Version} column also represents this distinction.
For each testing session, we report the number of runs of the transpiler (column \textit{Transpiler runs}), that is, the number of executed test cases, the number of unique injected faults evaluated at least once (column \textit{Injected faults}), as well as the number of exposed crashes (column \textit{Crashes}) and confirmed oracle failures  (column \textit{Oracle failures}).
%column indicates how many times the fault-injected transpiler has been executed on the follow-up programs to perform the translation process, without taking into account the confirmation process we previously outlined. The \textit{Injected Faults}
%column represents the total number of unique faults that have been injected into the transpiler. 

We observe that the number of unique injected faults is generally lower than the number of transpiler runs.
This is due to two primary causes. First, some transpiler runs may not traverse any yet-non-executed fault. Second, since transpiler runs are done in parallel, synchronization delays may lead them to activate the same injected fault.
%This is due to both some transpiler runs that may not traverse any yet-non-executed fault, and some transpiler runs done in parallel instances of the tools that occasionally may happen to execute the same injected fault due to synchronization delays.

%Indeed, this number does not match the run count due to the parallelism races introduced by our experimental setup, as the instrumentation tool synchronized the list of currently executed faults only at the beginning of the translation process.
Furthermore, no data on oracle failures is available for the fuzzer version of the tool, as the fuzzer does not check oracles.   
%The \textit{Detected via Crashes} and \textit{Detected via Oracles} column represent the number of injected faults that our techniques have been able to detect, respectively, due to the transpiler crashing and due to one of the consistency checkers of a metamorphicrelation reporting a violation. Naturally, this second column is only available for thenon-ablated version of our technique.

The results in the table provide empirical evidence of the merit of \acronym metamorphic testing, as the number of injected faults detected thanks to the metamorphic-testing oracles is significant, specially when compared to the number of faults that cause crashes. The failures reported based on the oracles correspond to cases in which the transpiler fails silently, as it indeed produces the output sources, even though the code in those sources is an incorrect translation of the code in the input sources. As a matter of facts, those failures cannot be detected with pure fuzzing: as the transpiler does not crash in those cases, a fuzzer can only observe that the output was indeed produced, being it agnostic about the actual code in those sources. 

Answering RQ1, we observe that in the testing sessions with \nbcc the number of failures detected with metamorphic oracles ranges between 6 and 43, with a median value of 39, whereas the number of crashes ranges between 5 and 16, with a median value of 13.

The testing sessions with the fuzzer version of the tool, where we ablated the use of the metamorphic  oracles, show that the overhead of \nbcc over a pure fuzzer (based on the same mutation operators) is negligible. In several cases the fuzzer version ends up with executing the same amount of transpiler runs as \nbcc, or just a few more. This can be explained because the application of the mutations and the execution of the transpiler dominate the execution time of the tool, while the evaluation of the oracles takes little time. If we consider the first testing session, the fuzzer version of our tool executed 1,952 transpiler runs, running four tool instances for 12 hours, which on average corresponds to generating and executing a test case every 89 seconds\footnote{With some particular DSL source inputs the transpiler may take even more than 10 minutes, as it happened for example in testing session 3.}, whereas the evaluation of the oracles takes hundredths of seconds. The comparison between the number of injected faults evaluated in the testing sessions with \nbcc and the corresponding testing sessions with the fuzzer version, respectively, and the comparison of the number of crashes observed in those testing sessions, respectively, confirm that there is no significant gain in dismissing the metamorphic oracles, conversely it pays the penalty of not exposing many failures.

%over pure fuzzing

%By comparison of the total number of detections with the number of injected faults, we notice that the detection rate is around $7\%$, with the highest amount obtained during the third run ($8.21\%$). Moreover, we notice how the number of injected faults our technique can detect thanks to the oracles is on average two and a half times higher than the one detected through crashes.

%Comparing \nabcc with its ablated version shows that the overhead of the oracles is quite small, as the total number of transpiler runs remains within the same order of magnitude, differing at most by $12$. 
%Run 5 and Run 7 represent interesting cases of this discrepancy due to the difference being skewed in favor of the non-ablated variant; a closer analysis of the results shows that this is due to slight deviations in timing that are compatible to OS-determined differences. Small differences can also be observed by comparing the number of injected faults or total failures detected through compiler crashes, with a respective maximum difference of $5$ and $2$ respectively.

In summary, answering \textbf{RQ1}, our case study provides empirical evidence that the \acronym approach to  metamorphic testing of transpilers is effective, as it is able to expose significant amounts of injected faults.  Answering \textbf{RQ2}, we observed negligible overhead between using or not using the metamorphic oracles, 
%is proven to be negligible 
as we can execute almost equal amounts of transpiler
runs and reveal comparable amounts of crashes
in both cases.

\subsection{Threats to Validity}
We could not compare \nbcc with baselines derived from existing fuzzers and existing approaches to metamorphic testing of compilers (such has the ones mentioned in section~\ref{existing:techniques}). Common fuzzers, like AFL, would generally fail in generating valid sources, while all existing approaches for metamorphic testing of compilers rely on executing the compiled programs, which is inapplicable in our case (as we discussed). 
%since they would not be applicable to our use case.
We mitigated this issue by creating a baseline that adapts \nbcc to work as a fuzzer, applying the same mutations used in \nbcc.
%, but without exploiting knowledge of the underlying metamorphic relations and oracles.

We do not claim that our results generalize beyond the specific transpiler and metamorphic relations considered in our experiment. 
%The same limitation applies to the number and type of metamorphic rules that are currently defined in \nbcc for our DSL. 
As future work we plan to increase the number and diversity of metamorphic relations considerd for the partner's DSL,
%future in agreement with our industrial partner and in the future we plan to 
evaluate the effectiveness of our technique for other transpilers. 

A threat to the validity of the results obtained in our experiment may come from the injected seeded faults that are only a subset of all possibly representative faults. While this is true, we affirm that our subset of injected faults is significant enough to show the benefits of our technique.

Internal validity may be threatened by the high number of random choices in our technique (spot selection, DSL selection) and empirical setup (fault seeding) in combination with a limited time budget. To mitigate this issue, we repeated our experiments 10 times demonstrating that our technique is consistently able to identify potential faults. %In the future we plan to setup an empirical experiment in which \nbcc is left running indefinitely and faults are reported whenever found.
%\luca{da discutere il punto 5)}
%\luca{1) 1 solo transpiler 2) set limitato di relazioni metamorfiche 3) mancanza di altre baseline (altre tecniche valide), ad esempio fuzzer (Eifel) e EMI che non funzionerebbero. 4) Injected fault sono solamente un SUBSet dei potenziali fault che riteniamo ugualmente significativo per i nostri scopi 5) abbiamo testato solamente un injected fault per un singolo caso di test. Magari altri casi di test lo rivelerebbero quindi la stima è conservativa.}

\section{Related Work}\label{sec:related}
In Section~\ref{sec:example}, we have already commented on why existing approaches to metamorphic testing of compilers~\cite{le2014compiler,le2015finding,tao2010automatic,nakamura2016random,donaldson2017automated} can hardly be directly applied for testing transpilers, such as the one used in our case study. Those approaches assume that the compiler under test produces executable programs for which inputs are readily available or easily derived, and their metamorphic relations assert that specific changes in the source programs should preserve the runtime behavior of the compiled programs.

In contrast, transpilers yield results that are themselves source code, which in many practical cases may depend on non trivial compilation and execution requirements. In our case study, the transpiler generates control‑logic programs that can run only when integrated with dedicated hardware or specialized simulators, and whose inputs depend on the structure and dynamics of the controlled plant. Under these conditions, the above-mentioned techniques for metamorphic testing of compilers are inapplicable. Our approach overcomes this limitation by defining metamorphic relations directly over the source code produced by the transpiler.

%In section~\ref{sec:example}, we have already commented on the hardness of reshaping existing approaches to metamorphic testing of compilers to address metamorphic testing of transpilers (as for instance the transpiler considered in the case study that we discussed in the previous section). The existing approaches assume that the compilers under test yield  compiled programs can be straightforwardly executed with available or easily derivable inputs, as they propose metamorphic relations predicating that given changes in the source programs shall result in compiled programs that behave equivalently at runtime~\cite{le2014compiler,le2015finding,tao2010automatic,nakamura2016random,donaldson2017automated}.
%However, the output of a transpiler is itself a source code, which in practical cases may be subjects to non trivial compilation and execution requirements. For instance the transpiler considered in our case study generates control logics, which can be executed only when integrated with dedicated hardware or suitable simulators, and whose inputs include the structure and the dynamics of the plant being controlled. In this scenario, the above-mentioned existing approaches to metamorphic testing of compilers are inapplicable. Our approach addresses this challenge by defining metamorphic relations that predicate on the source code yielded by the transpiler.  

Other compiler‑testing techniques rely on pure fuzzing~\cite{yang2011finding,sirer1999using,yoshikawa2003random,zhang2017skeletal,marcozzi2019compiler}, which is limited to revealing only observable crashes (as our case study confirms), or exploit differential testing~\cite{ofenbeck2016randir,sun2016finding2,hawblitzel2013will,li2023finding}, which is applicable only when multiple, supposedly equivalent implementations of the transpiler exist, a condition that is seldom met in practice and not satisfied in our setting.

Our approach defines metamorphic relations that ground on
mutating the input source code in the style of classic mutation analysis of programs~\cite{ma2006mujava,just2014major}.
Program mutations have been used by other authors in the literature on metamorphic testing. 
%to obtain programs that can then be tested according to metamorphic relations is not completely new, but existing techniques leverage this combination in a different way from ours.
For instance, some approaches automatically learn which mutants enable predefined metamorphic relations~\cite{nair2019leveraging}
or instantiate metamorphic relations by inferring how results change when applying given mutants~\cite{sun2016mumt}.
Some authors relied on mutation testing to evaluate the effectiveness of metamorphic relations~\cite{ayerdi2024genmorph,liu2012new,mayer2006empirical,sadi2011verification,saha2019fault}.
All these research efforts address different problems than transpiler testing.

\label{sec:rw}

\section{Conclusions}\label{sec:conclusions}

Motivated by the industrial relevance of transpilers, 
%due to the widespread adoption of DSLs, 
%automatic testing of transpiler aiming to validate their correctness remain unexplored.
this paper introduced a metamorphic testing technique, \nbcc, specifically designed to address automated testing of transpilers. \nbcc distinctively predicates metamorphic oracles over the source code that the transpiler under test produces as output, whereas existing approaches to metamorphic testing of compilers draw on executing the compiled binary. Our approach leverages on program mutations applied to the input sources provided to the transpiler. It grounds on metamorphic relations which specify how the mutations in the input sources shall consistently map onto corresponding equivalencies and differences in the output sources produced by the transpiler under test. 

We studied the effectiveness of \nbcc for testing the transpiler of an industrial partner, and used injected faults to assess the ability of \nbcc to reveal faults.
%assessed its ability by injecting it with seeded faults to assess the ability of \nbcc to reveal them.
The empirical results reported in the paper confirm that the proposed approach has the potential to 
reveal 
%indicate that our technique can effectively detect 
faults that would remain unexposed with pure fuzzing.
%, testifying the practical significance and the potential of the metamorphic technique presented in this paper.
We are currently working to extend our case study with metamorphic relations for all the constructs of the DSL of our partner, and to experiment the \acronym testing approach to other transpilers.
%defined in \nbcc and to plan an experiment in which \nbcc is left running indefinitely and faults are reported whenever found.

\textbf{Data Availability Statement}:
%Due to the proprietary nature of the DSL and transpiler that we used for our experiments, we are not able to provide the inputs and relative outputs to replicate our study. Instead, 
Our tool \nbcc is  open source and accessible anonymously on Zenodo at~\url{https://doi.org/10.5281/zenodo.19340835}. The same repository also includes the integral version of the metamorphic relations of Figure 3. For confidentiality reasons, we cannot disclose other data of the case study discussed in the paper, but the anonymized information  in Section~4.
%We provide our prototype \nbcc in which a transpiler to test and the associated metamorphic oracles need to be provided by the user, available at the URL~\url{https://anonymous.4open.science/r/ASE-McpTester}.

%\begin{acks}
%Acknowledgments 
%\end{acks}

\bibliographystyle{ACM-Reference-Format}
\bibliography{bib}

%\appendix


\end{document}